# Controlled assembly of graphene sheets and nanotubes: fabrication of suspended multi-element all-carbon vibrational structures

I. Tsioutsios,[1,2] J. Moser,[1,2] J. A. Plaza,[3] A. Bachtold [1,2,*]

1- ICFO, Av. Carl Friedrich Gauss, 08860 Castelldefels, Barcelona, Spain
2- ICN, CIN2-CSIC, Campus UAB, 08193 Barcelona, Spain
3- IMB-CNM (CSIC), E-08193 Bellaterra, Barcelona, Spain

We report on the fabrication and operation of a multi-element vibrational structure consisting of two graphene mechanical resonators coupled by a nanotube beam. The whole structure is suspended. Each graphene resonator is clamped by two metal electrodes. The structure is fabricated using a combination of electron-beam lithography and atomic-force microscopy nano-manipulation. This layout allows us to detect the mechanical vibrations electrically. The measured eigenmodes are localized in either one of the graphene resonators. The coupling due to the nanotube is studied by measuring the shift of the resonance frequency of one graphene resonator as a function of the vibration amplitude of the other resonator. Coupled graphene resonators hold promise for the study of nonlinear dynamics, the manipulation of mechanical states, and quantum non-demolition measurements.

*adrian.bachtold@icfo.es

## I. INTRODUCTION

Graphene sheets and carbon nanotubes can be used to fabricate mechanical resonators that possess a wide variety of outstanding properties.[1-8] These resonators can be employed as sensitive mass detectors,[9] their resonance frequency can exceed 10 GHz,[10,11] they exhibit strong mechanical nonlinearities,[7,12,13] and their mechanical vibrations can efficiently couple to electrons in the Coulomb blockade and the quantum Hall regimes.[3,4,14-16] These experiments have been carried out with resonators consisting of either one nanotube or one graphene sheet.

A natural extension of these works is to fabricate devices in which several nanotube and/or graphene mechanical resonators are coupled. A first step in this direction was made in an experiment where the coupling between two vibrating nanotubes was studied by gluing several nanotubes on a tip and by imaging them in a transmission electron microscope.[17] However, it is important to develop coupled vibrational structures with a well-defined layout in order to enable such experiments as the study of nonlinear dynamics, the manipulation of mechanical states, and quantum non-demolition measurements. The challenge presented by the fabrication of such structures is that nanotubes and graphene cannot be structured as easily as other materials. Indeed, graphene can be patterned into complicated shapes using electron-beam lithography and reactive-ion etching, but such graphene structures are fragile and often tear when suspended. As for nanotubes, they are difficult to bend into a controlled shape and to place at predetermined locations.

Here, we demonstrate the fabrication and the operation of a multi-element vibrational structure consisting of two graphene plates linked by a nanotube beam. The whole structure is suspended. The structure is fabricated using a combination of electron-beam lithography and atomic-force microscopy nano-manipulation. Each graphene plate is clamped by two metal electrodes, so that mechanical vibrations can be both actuated and detected electrically using the mixing technique.[2,18] Two mechanical eigenmodes are measured, each corresponding to vibrations localized in a different graphene plate. The coupling between the eigenmodes is evaluated by measuring the shift of the resonance frequency of one graphene plate as a function of the estimated vibration amplitude of the other plate.

**II. FABRICATION**

The fabrication process starts by depositing graphene flakes on highly doped, oxidized silicon wafers using the mechanical exfoliation technique (Figure 1(a)).[19] Single and bilayer graphene flakes are selected with an optical microscope by measuring the intensity of the reflected light (after calibrating the intensity with flakes characterized by Raman spectroscopy). The flake is patterned into two parallel rectangular plates using electron-beam lithography (EBL) and reactive ion etching in oxygen (Figure 1(b)). A dichloroethane solution containing multiwall carbon nanotubes is spin-cast onto the wafer (Figure 1(c)). The tip of an atomic force microscope (AFM) probe is then used to position the nanotube across the two graphene plates (Figure 1(d)).[20,21] Each graphene plate is contacted to a pair of Au/Cr electrodes by EBL, metal deposition and lift-off (Figure 1(e)). The graphene plates and the nanotube are suspended by etching 260 nm of the underlying silicon oxide in hydrofluoric acid and released in a critical point drier (Figure 1(f)). The highly doped silicon substrate is used as a backgate. Figure 1(g) shows a colorized scanning electron microscope image of a device made from a bilayer graphene sheet upon completion of the fabrication process.

Three operating devices were fabricated. We first present measurements obtained with one of them at a temperature of 4.2 K. The two graphene plates have the same length of 1.14 μm (between the clamping electrodes) and the same width of 1 μm. The length of the nanotube is 1.74 μm and its diameter is 17 nm. The electrical two-point resistances of the two graphene plates range from 40 to 100 kΩ. In comparison, the resistance of the multiwall carbon nanotube measured between the two graphene resonators is about 1 MΩ, and is therefore much larger. Thus, the electrical current flowing through the nanotube is negligible in the measurements discussed below.

## III. EXPERIMENTAL RESULTS

We first characterize our device by applying a driving force onto one of the two graphene plates. To do so, one graphene plate is actuated and its vibrations are detected using the frequency modulation (FM) mixing technique,[18] while the other graphene plate is kept electrically floating.[22] Figure 2(a) shows one prominent mechanical resonance in the spectrum of each individually driven graphene plate. Interestingly, the resonances appear at two distinct frequencies. This indicates that each graphene plate hosts one eigenmode; the vibrations of one graphene plate are transferred to the other plate only weakly.[23,24] In other words, the nanotube has a limited influence on the detected vibrational eigenmodes. Similar results are obtained with the two other measured devices. We note that, in Figure 2(a), the two graphene plates feature different current amplitudes on resonance as well as different background noises; this is attributed to the different electrical properties of the two graphene plates.[22]

The graphene plates are found to be under tensile stress by measuring their resonance frequency $f_0$ as a function of the constant voltage $V_{BG}$ applied to the backgate (Figure 2(b)). The resonance frequency decreases quadratically upon increasing $V_{BG}$. Similar results were obtained in previous measurements on single graphene resonators at low temperature.[6,7,25] The convex parabola has an electrostatic origin and indicates that the graphene plate is under tensile stress because of the metal electrodes, which contract upon lowering the temperature. The tension $T_0$ within each graphene plate can be quantified by fitting the $V_{BG}$ dependence of the resonance frequency to the expression derived for a resonator under tensile stress, $f_0(V_{BG}) = f_{max} - \sigma V_{BG}^2$ where $f_{max} = \frac{1}{2}\sqrt{T_0/mL}$ and $\sigma = f_{max} C''L / (4\pi^2 T_0)$. Here, $m$ is the mass of the resonator, $L$ its length, and $C''$ the second derivative of the graphene-gate capacitance with respect to displacement. The term $-\sigma V_{BG}^2$ originates from the plate oscillating in an electric field gradient. We find that the tension is 713 nN and 883 nN and the mass is 7.8 fg and 5.8 fg for graphene plates 1 and 2, respectively (see the plate labelling in Figure 2(a)). The corresponding mass densities are 9.2 and 6.9 times larger than that of pristine graphene, suggesting contamination of the graphene surface. Similar values of tensions and mass densities were found in previous measurements on single graphene resonators.[6,7,25] This further supports our finding that the nanotube affects the resonance of the graphene plates only weakly. The difference in mass density between the two graphene plates may be attributed to the partial cleaning of the contamination during the manipulation of the nanotube with the AFM tip during the fabrication of the device.[26]

Increasing the driving force applied to one of the graphene plates shifts its resonance frequency to higher values (Figure 2(c)). The resonance frequency is determined as the frequency for which the current measured with the FM mixing technique is largest[22] (we verified that this frequency is nearly equal to the frequency for which the motional amplitude is largest by solving the equation of motion numerically[7]). This behavior is attributed to the Duffing force that originates from the mechanical tension within the graphene plate at large motional amplitude; because it is clamped at both ends, the plate stretches and compresses periodically in time.

In order to gain insight into the vibrational properties of the device, it is useful to compare the masses and the spring constants of the nanotube and the two graphene plates. From the built-in tension estimated above, we derive the spring constants $k_{G1}$ = 6 N/m and $k_{G2}$ = 7.4 N/m for graphene plates 1 and 2, respectively. We calculate the mass and the spring constant of the nanotube by describing it as a doubly-clamped beam with no built-in tension, a good approximation for multi-wall nanotubes.[27] Using the mass density $\rho_{CNT} = 2200$ kg/m$^3$, the Young modulus E=0.3 TPa,[27] as well as the length (1.74 µm) and diameter (17 nm) of the nanotube measured with AFM, we derive a mass $m_{CNT}$ = 0.2 fg and a spring constant $k_{CNT}$ = 0.1 N/m for this nanotube. These values are much lower than those of the graphene plates. This again is in line with our finding that the nanotube should not strongly modify the mechanical eigenmodes of the graphene plates.

We now study the coupling between the two graphene plates. We realize this in a pump-probe experiment, as follows. Graphene plate 1 is probed by continuously recording its resonance lineshape with the FM technique while sweeping the frequency of the (pump) force applied to graphene plate 2 (Figure 3(b)). Figure 3(a) shows that the resonance frequency $f_1^0$ of plate 1 shifts when the pump frequency is swept through $f_2 \approx$ 180 MHz. This is the frequency at which plate 2 resonates. This shows that the resonance frequency of one plate depends on the motional amplitude of the other plate, which clearly demonstrates the existence of a coupling between the two eigenmodes of the system.[12,28-30] The asymmetric shape of the peak in $f_1^0$ as a function of $f_2$ is attributed to the Duffing force. In another device, the shift in $f_1^0$ is measured as a function of the amplitude of the pump force (Figure 4): it is consistent with the quadratic dependence expected from the theory of eigenmode coupling.[28-30]

We estimate that the strength of the eigenmode coupling of the first device is about $\approx$ 90 kHz/nm$^2$ using a shift in $f_1^0$ of 200 kHz (Figure 3(a)) and a motional amplitude $x_2$ of plate 2 of 1.5 nm. The latter is estimated in an approximate way, since we neglect the Duffing nonlinearity and use $x_2 = QC'V_2^{AC}V_{BG}/k_{G2}$ with the pump voltage $V_2^{AC} =$ 40 mV, $V_{BG} =$ 5.8 V, the derivative of the capacitance with respect to displacement $C' =$ 11 pF/m (estimated from the device geometry), and the measured quality factor $Q_2 =$ 4000. The eigenmode coupling of the second device is $\approx$ 60 kHz/nm$^2$ from Figure 4 ($x_2$ is estimated to be $\approx$ 1.6 nm for $V_2^{AC} =$ 50 mV).

## IV. DISCUSSION

The dynamics of two coupled mechanical resonators can be described by the set of equations

$$\ddot{x}_1 + \gamma_1 \dot{x}_1 + \omega_1 x_1 + \alpha_1 x_1^3 + D(x_1 - x_2) = F_1(t), \qquad (1)$$

$$\ddot{x}_2 + \gamma_2 \dot{x}_2 + \omega_2 x_2 + \alpha_2 x_2^3 + D(x_2 - x_1) = F_2(t), \qquad (2)$$

where $x_i$ is the displacement of the fundamental mode of resonator i, $\gamma_i$ its damping rate, $\omega_i$ its angular resonant frequency, $\alpha_i$ its Duffing coefficient, $F_i$ its driving force, and $D$ the coupling strength.[28] The coupling is attributed to the nanotube link (and not to the ledge of the metal electrodes,[28] since the graphene sheets are anchored to different electrodes). This coupling leads to two eigenmodes. Our measurements suggest that each eigenmode is essentially localized in a graphene resonator. Another consequence of the coupling is that the resonance frequency of one eigenmode depends quadratically on the motional amplitude of the other eigenmode,[28] which is consistent with our measurements. A quantitative estimate of this frequency dependence from Eqs. 1 and 2 requires a precise knowledge of the shape of the eigenmodes. We carried out simulations of our device with a finite-element method,[22] but the shape of the eigenmodes is very sensitive to various parameters that are unknown, such as the spatial distributions of the contamination and of the mechanical tension. A quantitative estimation of the coupling will necessitate further work, such as imaging the shape of the eigenmodes.[31]

## V. CONCLUSION

We have demonstrated the fabrication and operation of a multi-element vibrational structure composed of two graphene resonators coupled by a nanotube. Each measured eigenmodes is localized in a graphene plate. Because of this coupling, the motion of one graphene plate affects the motion of the other plate. Coupled resonators based on nanotube and graphene hold promise for the study of nonlinear dynamics, such as synchronization,[32] chaos,[28] Landau-Zener transition,[33] parametric mode splitting,[34] and the coherent manipulation of phonon population.[34,35] Indeed, owing to the low dimensionality of nanotube and graphene, mechanical nonlinearities emerge at relatively low driving forces and strongly affect their dynamics.[7,12,13,36,37] Coupled mechanical resonators also offer alternate strategies to improve the quality factor,[38] as well as to detect charge[39] and mass[40] with high sensitivity. In particular, it will be interesting for mass sensing to fabricate coupled resonators where the eigenmodes are delocalized over the two graphene sheets. Indeed, it has been shown that the amplitude of the eigenmodes is then extremely sensitive to the addition of mass onto the resonator.[23,24] In the quantum regime,[41-44] it has been proposed to use such nonlinear couplings between the eigenmodes for quantum nondemolition measurements.[45] In this context, an interesting feature of nanotube and graphene is that the amplitude of the zero-point motion is large.

## ACKNOWLEDGEMENTS

We thank M. Sledzinska and A. Eichler for experimental support. We acknowledge support from the European Union through the RODIN-FP7 project, the ERC-carbonNEMS project, and a Marie Curie grant (271938), the Spanish state (FIS2009-11284), the Catalan government (AGAUR, SGR), and by the MINAHE4 project (TEC2011-29140-C03-01).

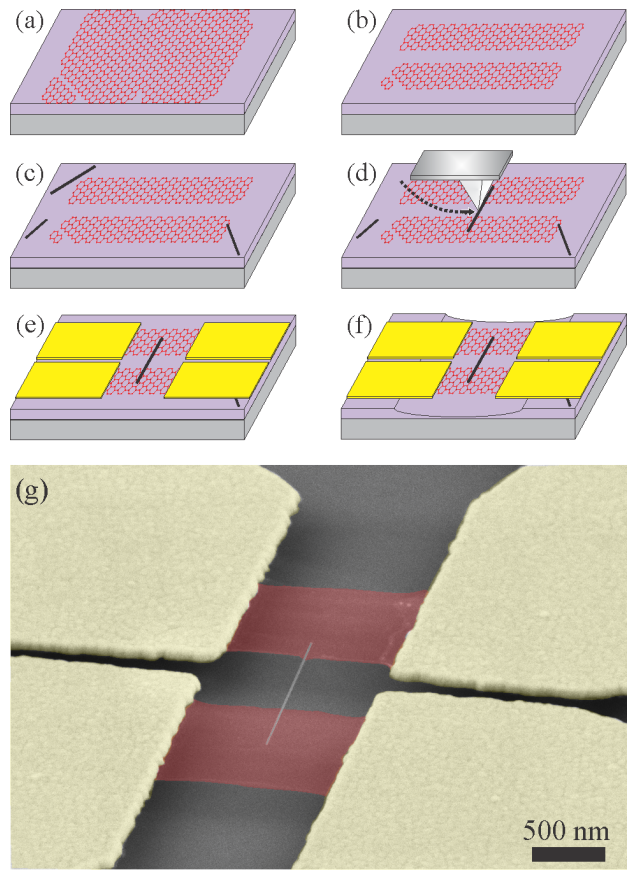

**FIG. 1.** Fabrication of two graphene resonators coupled by a nanotube. (a) Mechanical exfoliation of graphene onto an oxidized silicon wafer. (b) Shaping the graphene layer with reactive ion etching. (c) Deposition of nanotubes. (d) Manipulation of a nanotube with an AFM tip. (e) Patterning of metal electrodes using electron-beam lithography. (f) Removal of the silicon oxide below the structure with hydrofluoric acid. (g) Colorized scanning electron microscope image of the device at the end of the fabrication process.

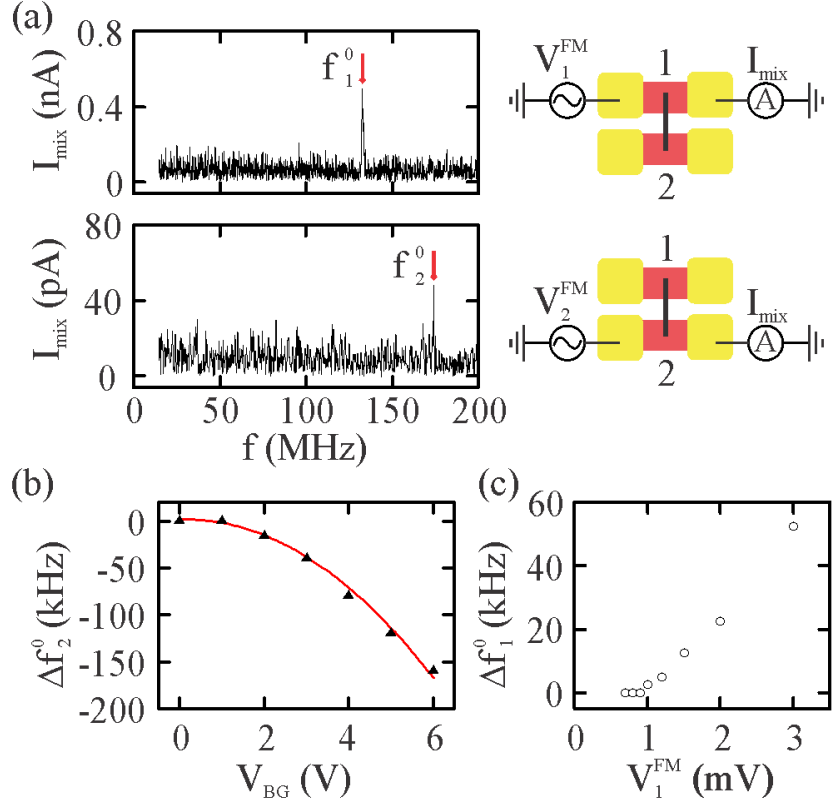

**FIG. 2.** Characterizing the coupled resonator made from a single layer graphene sheet. (a) Mechanical spectrum of graphene plates 1 and 2 (obtained by measuring the mixing current $I_{mix}$ as a function of the driving frequency). The schematics on the right-hand side show the measurement configuration. The gate voltage $V_{BG}$ is 5 V for the upper spectrum and -4 V for the lower spectrum. The quality factors are $Q_1 = 5500$ and $Q_2 = 4000$ for plates 1 and 2, respectively. $V_1^{FM}$ and $V_2^{FM}$ are the amplitudes of the FM oscillating voltages. (b) Resonance frequency shift of the second mode ($f_2^0$) as a function of $V_{BG}$. See Ref. 22 for the raw data. (c) Resonance frequency shift of the first mode ($f_1^0$) as a function of the driving force (proportional to $V_1^{FM}$).

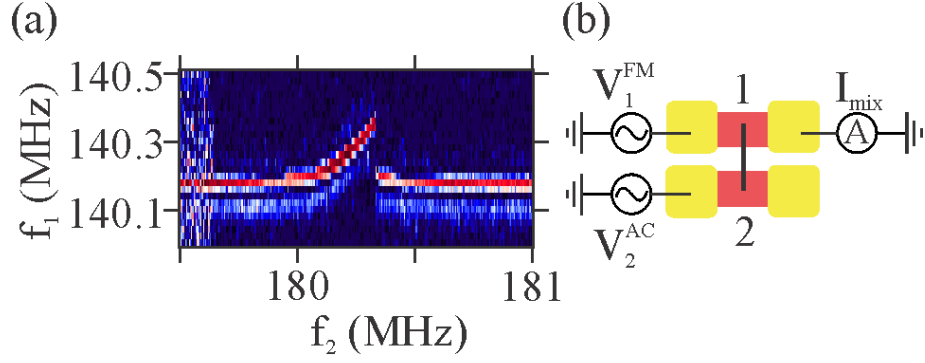

**FIG. 3.** Pump-probe experiment to study the coupling between the eigenmodes. (a) Resonance frequency of graphene plate 1 as a function of the frequency of the force applied to plate 2. The plot is obtained by continuously measuring the mixing current of plate 1 as a function of the frequency $f_1$ of the probe force, while sweeping the frequency $f_2$ of the pump force. The first mode is probed with $V_1^{FM} = 3$ mV, and the second mode is pumped with $V_2^{AC} = 40\ mV$. The gate voltage is 5.8 V. (b) Setup of the measurement scheme.

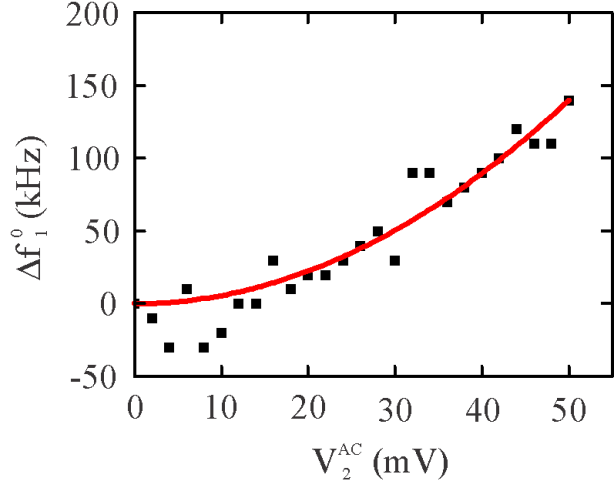

**FIG. 4.** Shift of the resonance frequency of plate 1 as a function of the pump voltage applied to plate 2. The measurement corresponds to a device made from a bilayer graphene sheet, which is different from the one discussed in the rest of the Letter. The first mode is probed with $V_1^{FM} = 5\ mV$. The gate voltage is 8 V. The resonance frequencies are $f_1^0 =$ 189.2 MHz and $f_2^0 =$ 175.2 MHz. The red curve corresponds to the quadratic dependence expected from the theory of eigenmode coupling.